\documentclass[aps,preprint,superscriptaddress]{revtex4}
\usepackage{graphics} % or graphicx
\usepackage{mathtools}
  \usepackage{lineno}
  \usepackage{cancel}
\usepackage{lipsum}
\begin{document}

%Title of paper
\title{Extended hybrid kinetic-magnetohydrodynamic model for ignited burning plasmas}
% \title{Steady state self-organized co-rotating dust vortices in complex plasma}
\author{Modhuchandra Laishram}\email{modhu@ustc.edu.cn}

\address{CAS Key Laboratory of Geospace Environment 
and Department of Engineering and Applied Physics, 
University of Science and Technology of China, Hefei 230026, China}
%  \address{Institute for Plasma Research, HBNI, Bhat, Gandhinagar 382428, India}

%   \author{2nd Author}
%   \author{Devendra Sharma}
% \address{Institute for Plasma Research, HBNI, Bhat, Gandhinagar 382428, India}

%  \author{3th Author}
\author{Ping Zhu}%\email{zhup@hust.edu.cn}

\address{International Joint Research Laboratory of 
Magnetic Confinement Fusion and Plasma Physics, 
State Key Laboratory of Advanced Electromagnetic Engineering and Technology, 
School of Electrical and Electronic Engineering, 
Huazhong University of Science and Technology, Wuhan, Hubei 430074, China}

\address{Department of Engineering Physics, University of Wisconsin-Madison, 
Madison, Wisconsin 53706, USA}
\author{Yawei Hou}
\address{CAS Key Laboratory of Geospace Environment and Department of Engineering 
and Applied Physics, University of Science and Technology of China, Hefei 230026,
China}
% % \author{3th author}
% \address{CAS Key Laboratory of Geospace Environment and Department of Engineering 
% and Applied Physics, University of Science and Technology of China, Hefei 230026,
% China}
% \affiliation
\date{\today}

\begin{abstract}
A brief review of the existing kinetic-magnetohydrodynamic(MHD) 
hybrid models for 
the alpha particle physics in burning plasma demonstrates that 
the pressure-coupling scheme is equivalent to the current-coupling 
scheme only in a specific dynamic regime where 
the alpha particle density is much lower 
than the background ion and electron. A more comprehensive 
kinetic-multifluid model is proposed for a proper account of the 
dynamical regime of the burning plasma where both the energetic alpha 
and the helium ash particles are present. 
The Kinetic-multifluid model is further 
simplified into an extended hybrid kinetic-MHD model in the MHD limit. 
This reduction process demonstrates that the existing 
pressure-coupling scheme is more extensive 
than the current-coupling scheme and sufficient for the wide range of 
dynamical regimes. This analysis further shows a significant 
change in the model equations mainly the generalized Ohm's law due to the 
contributions of a significant amount of alpha particles in the system. 
These models can be used for studies of the impact of the alpha particles 
present in ignited burning plasma and space plasma.
\end{abstract}
%%%%%%%%%%%%%%%%%%%%%%%%%%%%%%%%%%%%%%%%%%%%%%%%%%%%%%%%%%%%%%%%%%%%
\maketitle
%%%%%%%%%%%%%%%%%%%%%%%%%%%%%%%%%%%%%%%%%%%%%%%%%%%%%%%%%%%%%%%%%%%%%%%%%%%% 
\section{Introduction}\label{introduction}
Ignition in burning plasma is an important stage of thermonuclear fusion 
reaction in which the electrically charged $\alpha$ particles 
resulting from the fusion reaction can heat up the plasma 
and enable the reaction to achieve a self-sustained 
condition~\cite{0741-3335-45-5-312,Zweben_2000,0029-5515-35-2-514,
doi:10.1063/1.1690763}. Thus, the fusion reaction passes a 
balance condition of output alpha 
power $P_{\alpha}$ to the external input power $P_{in}$ (from neutral 
beam injection (NBI), resonance frequency (RF), and Ohmic heating), 
beyond which external heating is no more 
necessary~\cite{doi:10.13182/FST12-A13497,0029-5515-35-2-514,
doi:10.13182/FST06-A1222}. This balance condition 
corresponds to $P_{\alpha}/(P_{\alpha}+P_{in})\ge 1/2$ or $Q\ge 5$ in 
term of the power gain $Q=P_{out}/P_{in}$, where, 
$P_{out}=P_{\alpha}+P_{n}\approx 5P_{\alpha}$
~\cite{0741-3335-45-5-312,doi:10.1063/1.1690763,doi:10.13182/FST06-A1222}. 
It means, at the higher value of the power gain ($Q>5$), there would be  
a substantial amount of alpha particles (both energetic and thermalized) 
in addition to the thermalized electrons and ions in the core of 
fusion plasma. In fusion plasma, energetic alpha particles are usually 
produced by the neutral beam injection (NBI) and the radio frequency (RF) 
heating in addition to the main fusion reactions~\cite{doi:10.13182/FST12-A13497,
0029-5515-35-2-514,doi:10.1063/1.1461383}, whereas, the thermalized alpha 
particles arises due to frequent collisions with the thermalized 
bulk particles 
% because, the majority of these particles is concentrated in the 
% core with velocities intermediate in between the thermal electron and ion 
% speeds
~\cite{Zweben_2000,0029-5515-35-2-514,0741-3335-39-5A-025,
doi:10.1063/1.3587080}.
Experiments in Tokamak Fusion Test 
Reactor (TFTR), Joint European Torus (JET), and JT-60U have confirmed 
the alpha particle generation even though it could not reach the state of 
self-sustained nuclear reactions~\cite{Zweben_2000,
doi:10.1063/1.1690763,0741-3335-45-5-312,doi:10.13182/FST06-A1222}.
Due to the lack of adequate datas, there has been very less 
discussion on the effects of the alpha particles and 
alpha-particle physics remain one of the active research 
area for fusion plasma~\cite{freidberg_2007}. 

To give an estimate on the $n_{\alpha}/n_b$, (i.e., the density ratio of 
alpha particles to the background plasma) for the existing and future 
tokamak experiments, we take the recent design of
CFETR (China Fusion Engineering Test 
Reactor)\cite{Chen2017,Shi2016,Yang2017}, which is a 
little bigger than ITER (International Tokamak Experimental 
Reactor)~\cite{BeyondITER}, as an illustration. 
The recent CFETR is designed with $7.2~m$ major 
radius, $2.2~m$ minor radius, $P_{out}\sim1~GW$ fusion power and 
$\beta_{\alpha}/\beta_b=0.1$~\cite{Wan2017,Chan2015}. 
Here $\beta_j=2\mu_0 p_j/{\bf B}^2$ is the 
ratio of thermal to magnetic energy of $j$-species~\cite{freidberg_2007}.
Considering the plasma pressure is proportional to density and 
temperature, the density ratio between $\alpha$ particles and background 
plasma should be $n_{\alpha}/n_b=(\beta_{\alpha} T_b)/(\beta_b T_{\alpha})$.
In a burning plasma, $\alpha$ particles are a mixture of 
thermalized helium ash (of density $n_0$)
and slowing down particles (of density $n_f$)
i.e., $n_{\alpha}= (n_0 +n_f)$. Here, the subscripts $0$ and $f$ denote the 
thermalized helium ash and energetic alpha particles respectively.
The character temperature of 
background plasma, slowing down, and helium ash are picked 
to be $30~KeV$, $1.5~MeV$ and $90~KeV$, respectively. Therefore, the approximate 
density ratio $n_{\alpha}/n_b$ would be $0.002$ and $0.033$ for slowing 
down $\alpha$ particles and helium ash, separately. 

Further, from the relations, 
$P_{out}=n_D n_T <\sigma v> \epsilon=n_b^2 <\sigma v> \epsilon$ 
and $P_{out}=5P_{\alpha}=5n_{\alpha}\epsilon$, fusion power $P_{out}$ is 
proportional to $n_b^2$ and $n_{\alpha}$ separately~\cite{freidberg_2007}. 
Where, $P_{out}$ is fusion power, $P_{\alpha}$ is the 
power of $\alpha$ particles, 
$n_D$, $n_T$, and $n_b$ are deuterium, tritium 
and background plasma, resistivity. $\sigma$ is the cross-section of 
deuterium and tritium, $v$ is the velocity of particle and $\epsilon$ is the 
energy released per reaction. Then, density ratio $n_{\alpha}/n_b$ is 
proportional to $\sqrt{P_{out}}$. If $P_{out}$ is increased to 
be $10GW$, $n_{\alpha}/n_b$ should be $0.007$ and $0.11$ for slowing 
down $\alpha$ particles and helium ash, separately. If $P_{out}$ is increased 
to be $100HGW$, $n_{\alpha}/n_b$ should 
be $0.02$ and $0.33$ for slowing down $\alpha$ particle and helium ash, 
separately. Although the slow down $\alpha$ particles density is less 
$n_f\ll n_b$, the helium ash density is $n_0\simeq n_b$ ($\sim 10\%$), 
thus need to modeled as differently than the slow down alpha 
particles. Systematic studies of the effect of both energetic 
$\alpha$ particles and thermalized helium ash on the 
plasma behavior on and after the self-sustained condition are indeed 
very important and always necessary to control the thermonuclear reactions 
in future fusion devices~\cite{0741-3335-45-5-312,doi:10.1029/1998JA900065,
doi:10.1029/91JA01981}. 

% For the recent tokamaks, including ITER and CFETR, the 
% assumption $n_{\alpha} \ll n_b$ is still valid. 
% In other words, when $P_f$ is bigger than $10GW$, we can consider 
% the $\alpha$ particle density effect.

% At the moment, there have been many efforts to study the physics 
% of energetic alpha particles physics in burning plasma 
% by extending well known MHD formulation both analytically and numerically.
Hybrid kinetic-MHD models such as pressure-coupling 
model~\cite{KIM2004448,doi:10.1063/1.4999619,doi:10.1063/1.3587080} and 
current-coupling model~\cite{doi:10.1063/1.860011,
doi:10.1063/1.4774410,doi:10.1063/1.872791,BELOVA1997324,Todo_2016} 
are two existing and equivalent models that have been using for 
studies of energetic $\alpha$ particle physics in 
burning plasmas. Thus, the model has been used 
successfully by several groups to study Alfven eigenmodes (AEs), 
Toroidal Alfven eigenmodes(TAE) modes, fishbone oscillations, and 
stability analysis of the modes that can be driven unstable by 
interaction with the energetic 
particles~\cite{doi:10.1063/1.872997,doi:10.1063/1.4999619,
doi:10.1063/1.1461383,BELOVA1997324,Todo_2016,doi:10.1063/1.871071}.
The equivalent condition and the whole derivation of the  
existing models are started under the assumption that the alpha particles 
are energetic and very rarefied i.e., $n_{0}\approx 0$, 
$n_\alpha = (n_0 +n_f)\approx n_{f}$, 
and $n_{f}\ll n_b$~\cite{doi:10.1063/1.873437,doi:10.1063/1.3587080}.  
%
% Therefore, these existing models are 
% sometimes limited depending on the amount of 
% approximations considered in the formulation and their practical 
% applicability
% However, there are many unsolved issues that involve the 
However, in the ignited burning plasmas~\cite{0029-5515-35-2-514,
doi:10.1063/1.1461383}, the thermalized helium ash density is sometimes 
comparable of background plasma density i.e., 
$n_0\gg n_f$ and $n_{\alpha} \approx n_b$. 
A sufficiently large population of alpha particles both energetic and 
thermalized helium ash can significantly influence plasma equilibrium, 
stability, transport, and the confinement
~\cite{doi:10.1063/1.2949704,freidberg_2007,
doi:10.1063/1.4908551,doi:10.1063/1.4774410}. 
% drive 
% instabilities, destabilize the existing modes, and 
% affects the whole plasma behaviors such as the pressure balance 
% for equilibrium, the empirical scaling relations for 
% energy confinement time, the bootstrap current, the resistivity and 
% many more
% Thus, the present scenarios of research on alpha particles and their 
% requirements in fusion plasma motivate us to 
Thus, studies of alpha particle plasma physics is one of the 
crucial physics goals of the next generation fusion plasma.

In this work, we review the limitations of the existing hybrid 
kinetic-MHD model and extend it to a more generalized 
kinetic-multifluid model to account for both the 
thermalized and the energetic alpha 
particles present in the burning plasma. 
% For this, we first check the 
% limitations of the existing models and then formulate a new 
% kinetic-multifluid model, in the presence of 
% both thermalized helium ash and non-thermalized energetic alpha 
% particles. 
In particular, the non-thermalized particles are modeled 
using the kinetic approach, whereas, the fluid model is
considered for the other thermalized particles. 
% And, for the non-thermalized 
% energetic particles of distribution function $f_f(r_f,{\bf v}_f)$, it is 
% possible to approximate new macroscopic dynamic variables $T_f$, other than the 
% dynamical variables ($n_f,~{\bf v}_f,~{\bf P}_f$), which takes similar role of 
% temperature in energy equation, and coupling with other thermalized 
% species~\cite{10007639755,Tronci_2014}. The major difficulty of this 
% model is the governing equations which describe explicitly the role of 
% each dynamical variables, sources terms, and parameters 
% the generation or annihilation, the exchange between the multi-species,
% the thermal conductivity, the resistivity, and many more 
% which are contributing at varying scales, are all significant in the 
% model~\cite{doi:10.1029/1998JA900065,
% doi:10.1029/91JA01981,doi:10.1063/1.3587080,doi:10.1063/1.4994073}.
% In addition, the generation or 
% annihilation, the exchange between the multi-species, and other 
% sources for particle, momentum, and energy, which are contributing to varying 
% scales, are all significant in the model~\cite{doi:10.1029/1998JA900065,
% doi:10.1029/91JA01981,doi:10.1063/1.3587080,doi:10.1063/1.4994073}. 
% Therefore, this model will be appropriate for studies of non-ideal effects and 
% phenomena associated with frequencies even close to the ion 
% cyclotron frequency, and length scale close to the Larmor 
% radius~\cite{doi:10.13182/FST12-A13497,doi:10.1029/1998JA900065,
% doi:10.1029/91JA01981}. 
 %
Taking advantage of the MHD limit, 
the generalized kinetic-multifluid model can be reduced to a new 
extended hybrid kinetic-MHD model. This reduction to the single fluid model 
demonstrates a significant change in the resulting 
generalized Ohm's law due to the presence of a substantial amount of alpha 
particles in the system.~\cite{doi:10.1063/1.4890955}. 
% Interestingly, it could demonstrate 
% the extensive nature of Pressure-coupling scheme over the Current-coupling 
% scheme for $n_{\alpha}= (n_0 +n_f) \approx n_b$, $n_0\gg n_f$ case where 
% alpha particles play a significant role in the system. This model is 
% relatively simplified, and therefore, restricted only for describing 
% the low-frequency MHD waves and global transport phenomena in burning 
% plasma~\cite{doi:10.1063/1.873437,KIM2004448}. 
% Both the models will be useful for studies the various impact of both minor 
% energetic particles and helium ash present in the fusion plasma. 
% Interestingly, other than the burning 
% plasma, the interaction of energetic alpha particles with other low 
% temperature thermalized species take place in space plasmas such as the 
% energetic solar wind interacts with the magnetosphere, the ionosphere, and 
% the thermosphere~\cite{0741-3335-56-9-095008}. 
% Therefore, these models can also be 
% used for studies of multi-species charged system in space plasma.

Thus, the manuscript is organized as follows. In Sec.~\ref{Existing_model}, 
a brief review of the existing Kinetic-MHD Hybrid Models for burning 
plasma are analyzed regarding its limitations. Then, 
in Sec.~\ref{multi_formulation}, we present a detailed description 
of a more general kinetic-multifluid model in presence of comparable 
amount alpha particles to the background plasma. This is followed in 
Sec.~\ref{extended_MHD_formulation}, by a reduction 
to an extended hybrid kinetic-MHD model showing significant
changes in the generalized Ohm's law due to 
the presence of alpha particles. Summary and conclusions are 
presented in Sec.~\ref{conclusion}. 
This paper contains an appendix~\ref{appendixA}, 
which is included for a full derivation of the 
extended generalized Ohm's law.
%%%%%%%%%%%%%%%%%%%%%%%%

%%%%%%%%%%%%%%%%%%%%%%%%
\section{Existing Kinetic-MHD Hybrid Model}
\label{Existing_model}
%%%%%%%%%%%%%%%%%%%%%%%%%%%%%%%%%%%%%%%%%
Before introducing a  generalized model for $\alpha$ particles 
physics in burning plasma, we start from a brief review of the 
existing kinetic-MHD hybrid models and their limitations. 
The existing models assume that the plasma is consists of two main 
components, the low-density energetic $\alpha$ particles and the 
high-density bulk plasma components, 
i.e., $n_{\alpha}\ll n_b$ and $\beta_{\alpha}\approx \beta_b$. 
Besides, there is no 
consideration of thermalized helium ash in the existing models.
% In all the models, the energetic 
% components are treated by a kinetic approach, whereas the usual 
% MHD description for the main plasma 
% component~\cite{doi:10.1063/1.860011,KIM2004448,doi:10.1029/1998JA900065,
% Todo_2016}. Thus, we have 
Thus, in the pressure-coupling 
model~\cite{KIM2004448,doi:10.1063/1.4999619}, where the $\alpha$ particle 
physics is coupled to the bulk plasma equation through pressure tensor 
as follows,
% \begin{eqnarray}
\begin{align}
 \frac{\partial n_b}{\partial t} + \nabla\cdot(n_b{\bf v}_b)&=0, 
%  =\pm \gamma_i n_i \pm\sum_{j}^{} \dot{S}_{ij} 
\label{continuity}\\
%  \end{equation}
% \begin{equation}
n_bm_b\left(\frac{\partial {\bf v}_b}{\partial t}
+{\bf v}_b\cdot\nabla{\bf v}_b\right) 
&= {\bf J}_b\times{\bf B}-\nabla p_b-\nabla\cdot{\bf P}_{\alpha},
\label{pressue_coup} \\
% \end{eqnarray}
% \begin{align}
%\begin{split}
\frac{1}{\gamma_b-1}\left(\frac{\partial p_b}{\partial t} +
  {\bf v}_b\cdot\nabla p_b\right) &=-p_b\nabla\cdot{\bf v}_b ,
  \label{pressure_eqn} \\
%\end{split}
\frac{\partial {\bf B}}{\partial t} &= -\nabla\times {\bf E} ,\\
{\bf J}_b &= \frac{1}{\mu_0}\nabla\times{\bf B} ,\\
{\bf E} + {\bf v}_b \times{\bf B} &= 0.\label{ohms1_eqn}% \eta {\bf J}_b,
\end{align}
Where, the subscripts $b$ and ${\alpha}$ denote the bulk plasma and the 
${\alpha}$ particles, respectively. $n_b$, ${\bf v}_b$, and ${\bf J}_b$ are 
the particles density, the mean velocity, and the current of the bulk 
plasma, respectively. $p_b$ is the pressure affiliated with the 
thermalized bulk plasma, and ${\bf P}_{\alpha}$ is the pressure tensor 
associated with the energetic ${\alpha}$ particles which can be calculated 
using $\delta f$ method in the kinetic 
approach~\cite{KIM2004448,doi:10.1063/1.2949704}.
% Then, the associated 
% adiabatic, Maxwell, and Ohm's law equations which make the 
% fluid equations completely closed and also coupled with the 
% self-consistent electromagnetic fields are, 
$\gamma_b$ is a ratio of specific 
heats $(C_p/C_v)$, whose value depends on the nature of heat flux 
and its distribution~\cite{freidberg_2007}. ${\bf E}$ and ${\bf B}$ 
are self-consistent electric and magnetic field associated with 
the system. %Other notations are all conventional. 

In the current-coupling model~\cite{doi:10.1063/1.860011,
doi:10.1063/1.4774410,doi:10.1063/1.872791,BELOVA1997324}, 
the ${\alpha}$ particles charge and current are coupled 
to the bulk plasma momentum equation as follows,
 \begin{equation}
n_bm_b\left(\frac{\partial {\bf v}_b}{\partial t}
+{\bf v}_b\cdot\nabla{\bf v}_b\right) 
=({\bf J}_b-{\bf J}_{\alpha})\times{\bf B}
-q_{\alpha} n_{\alpha}{\bf E}-\nabla p_b. 
\label{current_coup} 
\end{equation} 

Where the rest of the equations remain the same.
Both the coupling models have been used 
successfully by several groups, and  generally adequate for describing 
the low frequency global behaviors and geometrical effects 
of the burning plasma that can be affected by interaction with the 
energetic particles~\cite{doi:10.1063/1.872997,doi:10.1063/1.4999619,
doi:10.1063/1.1461383,BELOVA1997324,Todo_2016,doi:10.1063/1.871071}.

Now, in general, one can define 
the macroscopic dynamical variables for ${\alpha}$ particles as the 
moments of corresponding distribution function 
$f_\alpha({\bf r}_\alpha, {\bf u}_\alpha)$ (from the probabilistic kinematic 
treatment~\cite{KIM2004448,doi:10.1063/1.2949704},) as 
follows,
\begin{eqnarray}
  {n}_\alpha &=& \int f_\alpha({\bf r}_\alpha,{\bf u}_\alpha) d^3 u_\alpha, \\
  {\bf v}_\alpha &=& \frac{1}{{n}_\alpha}\int{\bf u}_\alpha 
  f_\alpha({\bf r}_\alpha,{\bf u}_\alpha) d^3 u_\alpha, \\
  {\bf P}_\alpha &=& m_\alpha \int ({\bf u}_\alpha-{\bf v}_\alpha)
  ({\bf u}_\alpha-{\bf v}_\alpha)
  f_\alpha({\bf r}_\alpha,{\bf u}_\alpha) d^3 u_\alpha. 
  \label{eq:fmom}
\end{eqnarray}
The dynamics of the ${\alpha}$ particles can be 
approximated from the moment of Boltzmann kinetic 
equation such as,
 \begin{equation}
  m_\alpha\frac{\partial{n_\alpha\bf v}_\alpha}{\partial t} 
= q_\alpha n_\alpha{\bf E} + {\bf J}_\alpha\times{\bf B} 
-\nabla\cdot{\bf P_\alpha}.
\label{alpha_comp} 
\end{equation}
This equation clearly shows that the pressure-coupling model in 
Eq.~(\ref{pressue_coup}) is equivalent to 
the current-coupling model in Eq.~(\ref{current_coup}),  
only when the inertial term 
$m_\alpha\frac{\partial{n_\alpha\bf v}_\alpha}{\partial t}$ 
in Eq.~(\ref{alpha_comp}) is zero or negligible. 
Therefore, the equivalent condition of these existing hybrid models are 
restricted only to the regime where energetic particle density is 
relatively small ($n_{\alpha}\ll n_b$). 
Besides, there is no consideration of thermalized helium ash 
(of density $n_0$) in the existing models. 
% Thus, the inertial effect, the role of parallel electric field 
% in the Eq.~(\ref{alpha_comp}) are ignored, but their thermal 
% content is included through the pressure or the current only. 
Further, the existing models have been 
deficient in other aspects regarding the validation of the adiabatic 
energy equation and the ideal Ohm's law in the presence of 
${\alpha}$ particles~\cite{doi:10.1029/1998JA900065}.
The adiabatic or the equation of state given in 
Eq.~(\ref{pressure_eqn}) is valid only when the heat flow 
is negligible, and the dissipative process due to resistivity becomes 
important in the system with a substantial amount of ${\alpha}$ 
particles in burning plasma. 
Thus, a brief review of the existing kinetic-MHD hybrid models 
has demonstrated that the existing hybrid models are deficient 
in many aspects and require more generalization including the dynamic 
regime of $n_{\alpha}\approx n_b$, which would be observed on or after 
the ignited burning plasma. 
% Finding the appropriate formulation 
% for $n_{\alpha}\approx n_b$ is the main goal of this work. 
Therefore, in the following section, we propose a 
kinetic-multifluid model for proper treatment of the 
dynamical regime of burning plasma when 
$n_{\alpha}\approx n_b$. 
% regime, and reduces it again 
% to an approximate single fluid Extended Kinetic-MHD 
% Hybrid Model.

\section{Description of kinetic-multifluid model}
\label{multi_formulation}
A complete treatment for burning plasma carrying a substantial 
amount of alpha particles both thermalized helium ash and 
energetic particles would be a multifluid model that allows 
us to see physics of varying length scales and 
time scales~\cite{doi:10.1029/1998JA900065,
doi:10.1029/91JA01981,
doi:10.1063/1.3587080,doi:10.1063/1.4994073}. Thus, 
we introduce a more comprehensive kinetic-multifluid model 
for describing its various constituents 
such as the thermalized helium ash (of density $n_0$) and 
the energetic particles (of density $n_f$) in addition to the 
bulk plasma electrons (of density $n_e$) and ions ( of density $n_i$). 
For such a multifluid system having substantial amount 
of alpha particles %$n_{\alpha}(=n_f+n_0)\approx n_i$,
i.e., $n_{\alpha}= (n_0 +n_f) \approx n_i$, 
with possible cases of density $n_0\ll n_f$, $n_0\approx n_f$, 
and $n_0\gg n_f$,
% i.e., $n_{\alpha}= (n_0 +n_f) \approx n_i$ and $n_0\gg n_f$,
one can starts from the distribution function for each species, then 
define the macroscopic dynamical variables and moment equations 
respectively. 
Thus, starting from the zeroth moment of the Boltzmann 
equation, i.e., the continuity equations for each species of the 
multifluid system can be written as,
\begin{equation}
 \frac{\partial n_j}{\partial t} + \nabla\cdot(n_j{\bf v}_j) 
 = \eta_j n_j.  
\label{bulk_plasma_continuity}
 \end{equation}
Where, ${n}_{j}$ and ${\bf v}_{j}$ are 
average particle density and velocity of $j$th-species. 
$\eta_j$ in the source term is the generation or annihilation rate of 
each species which depends on the various system properties, and 
therefore, require 
a systematic analysis~\cite{doi:10.1063/1.4931169,freidberg_2007}. 
For example, ${\eta}_{j}$ is a function of fueling or 
reaction between a $j$th-species and other $k$th-species 
(electrons, ions, and alpha particles). 
And it could be zero for electrons and ions, whereas, 
$\eta_\alpha$ (for alpha species) is 
non-zero and function of space and time. %in burning plasma. 

The momentum balance 
equation (the first moment) of each $j$th-species can be written as, 
\begin{eqnarray}\nonumber
n_jm_j\frac{d {\bf v}_j}{d t}
=\eta_jn_jm_j{\bf v}_j -\nabla{p_j} +q_j n_j({\bf E}
+ {\bf v_j}\times{\bf B})\\
% \pm \gamma_j n_jm_j{\bf v}_j \\ 
+\sum_{k}^{} m_j n_j\nu_{jk}({\bf v_j}-{\bf v_k}),~~~~
% \pm m_i n_i\nu_{ik}({\bf v_i}-{\bf v_k}) 
% \pm m_i n_i\nu_{if}({\bf v_i}-{\bf v_f}) 
\label{bulk_plasma_momentum} 
\end{eqnarray}
where ${m}_{j}$ is mass, ${p}_{j}$ is the pressure 
of each $j$th-species, $\nu_{jk}$ is Coulomb interaction 
coefficient between the charged $j$th and $k$th-species. 
Now, the above momentum 
Eq.~(\ref{bulk_plasma_momentum})  
is meant for electrons, ions, and the thermalized helium 
ash (of the density $n_0$), but not for the non-thermalized 
energetic alpha particles of density $n_f(\approx n_\alpha-n_0)$. 
Therefore, for the energetic alpha 
particles with a distribution function 
$ f_f({\bf r}_f,{\bf u}_f)$, one can obtain the macroscopic 
dynamical variables ($n_f,~{\bf v}_f,~{\bf P}_f$)~\cite{KIM2004448}, 
and the corresponding momentum balance 
equation for the energetic alpha particle species can be approximated 
as the above Eq.~(\ref{alpha_comp}) (given in section~\ref{Existing_model}), 
with additional source term $\eta_fn_fm_f{\bf v}_f$ and Coulomb interaction 
terms ($ m_f n_f\nu_{fk}({\bf v}_f-{\bf v}_k)$) with other 
$k$th-species~\cite{doi:10.1063/1.860011}.

The energy equation (the second moment) for each thermal $j$th-species 
can be written as, 
\begin{eqnarray}\nonumber
\frac{1}{\gamma_j-1}\left(\frac{\partial p_j}{\partial t} +
  {\bf v}_j\cdot\nabla p_j\right) =-p_j\nabla\cdot{\bf v}_j
  -\nabla\cdot {\bf q}_j \\
  + S_j+\sum_{k}^{} 2 n_jm_j\frac{m_j}{m_k}\nu_{jk}({T_j}-{T_k}).  
\label{bulk_plasma_energy} 
\end{eqnarray}
Where, $\frac{p_j}{\rho_j(\gamma_j-1)}$ is the internal energy, 
$\gamma_j$ is the ratio of specific 
heats, ${\bf q}_j$ is the heat flux due to thermal conduction, $S_j$ is the 
sources of internal energy such as 
the ohmic heating, external auxiliary heating, radiation 
losses, and many more~\cite{freidberg_2007}. 
The last term represents the energy transfer rate 
due to the Coulomb interaction with other 
$k$th-species~\cite{doi:10.1063/1.873437}.
The temperature $T_j$ of each thermalized $j$th-species are well known and 
satisfies the relation $p_j=n_jT_j$, here, the unit of $T_j$ is chosen to make 
the Boltzmann's constant unity. 
However, the temperature for non-thermalized particles present in the 
system is not yet defined. For the non-thermalized 
state of the energetic alpha particles, we can approximate a new 
dynamical variable ${T}_f$ using the 
following, 
\begin{eqnarray}
{T}_f = (m_f/3n_f)\int |({\bf u}_f-{\bf v}_f)|^2
f_f({\bf r}_f,{\bf u}_f) d^3 u_f, 
\label{eq:fmom}
\end{eqnarray}
% Here, ${T}_f$ is function of space, time, and other 
% macroscopic characteristics of the non-thermalized species. 
% It also tends
which reduces to the roles of temperature in the case of thermalized 
species~\cite{10007639755,Tronci_2014}. 
Thus, the same form of approximate energy equation can be written 
for the variable ${T}_f$ along with the diagonal elements of the 
pressure tensor ${\bf P}_f$ of the energetic alpha particles 
present in the multi-species system. 
% However, in general cases, 
% where the ionizations, the reaction between the multiple species 
% is significant, and heat flow is not small, the different 
% species can't treat as the adiabatic as in the previous 
% treatments. 
% Now, each species of the complex multi-fluid system with enormous 
% amount of energetic particles can't be treated as adiabatic 
% as in the previous treatment. 
% However, we can define the corresponding energy equations for each 
% species under certain valid approximations. 

Furthermore, the charged multifluid system couple to the 
self-consistent electromagnetic fields through the 
current density $J$ and the charge density $\rho_q$ in the 
following Maxwell's equations,
\begin{eqnarray}
\nabla\cdot {\bf E} &=&\frac{\rho_q}{\epsilon_0},\\
\nabla\cdot {\bf B} &=&0, \\
\nabla\times{\bf B}&=& \mu_0{\bf J}
+\mu_0\epsilon_0\frac{\partial {\bf E}}{\partial t},\\
\frac{\partial {\bf B}}{\partial t} &=& -\nabla\times {\bf E}.
\label{bulk_Maxwell}
% {\bf E} + {\bf V}\times{\bf B} &=  \eta {\bf J},
% {\bf V'} &=& \frac{q_i n_i {\bf v}_i+ q_f n_f {\bf u}_f 
% + q_0 n_0 {\bf u}_0}{q_i n_i+ q_f n_f + q_0 n_0}\\ \nonumber
% {\bf E} + {\bf V'} \times{\bf B} &=& \eta {\bf J} ,\\ \nonumber
\end{eqnarray}
 Where,
\begin{eqnarray}
\rho_q = \sum_{j}^{} q_j n_j,~~and~~~
% \rho_m &=& n_e m_e + n_i m_i+ n_f m_f + n_0 m_0 \\
{\bf J} = \sum_{j}^{}q_j n_j {\bf v}_j.
\label{bulk_nutrality}
\end{eqnarray}
These sets of equations from Eqns.~(\ref{bulk_plasma_continuity})
to~(\ref{bulk_nutrality}) are the kinetic-multifluid model for 
a wide range of dynamical regimes of the burning plasma. 
This model will allow us to describe multi-scale phenomena in tokamak 
% such as finite Larmor radius effects, micro-turbulence, 
% thermal conductivity, resistivity effects, 
in addition to the 
studies of macroscopic equilibrium, instabilities, and transports 
processes %associated with the varying time scales and length scales in the 
~\cite{doi:10.1063/1.4890955,10007639755,freidberg_2007,10007639755}. 
% and many 
% more~\cite{doi:10.1063/1.4890955,10007639755,freidberg_2007}. 
% This model is also more accurate for studies of macroscopic 
% equilibrium, transports, and instability analysis associated with 
% the varying time scales and length scales in the 
% tokamak~\cite{10007639755}. 
% Interestingly, 
This model is also appropriate for 
study of multifluid phenomena in space plasma 
such as the coupling between the energetic solar wind and 
magnetosphere, ionosphere, and thermosphere~\cite{SONG2005447}.
% However, the model is full of complicated nonlinear terms 
% with the varying scales and so difficult to solve 
% simultaneously. 
The major difficulty of this 
model is the governing equations which describe explicitly the role of 
each dynamical variables, sources terms, and parameters such as  
the generation or annihilation, the exchange between the multi-species,
the thermal conductivity, the resistivity, and many more 
which are contributing to varying scales, are all significant in the 
model~\cite{doi:10.1029/1998JA900065,
doi:10.1029/91JA01981,doi:10.1063/1.3587080,doi:10.1063/1.4994073}.
However, when the real macroscopic effects are important, 
the above equation can be simplified into various form of MHD model. 
In the following section, we reduce 
the generalized kinetic-multifluid model to an extended hybrid 
kinetic-MHD model which can be useful for describing low frequency 
burning plasma processes in the presence of both 
% global behaviors and 
% geometrical effects of burning plasma that has enormous amount of 
thermalized and non-thermalized energetic alpha particles.  

\section{Reduction of kinetic-multifluid model to 
Extended hybrid kinetic-MHD model}
\label{extended_MHD_formulation}
%%%%%%%%%%%%%%%%%%%%%%%%%%%%%%%%%%%%%%%%%%%%%%%%%%%%%%%%%
Single-fluid MHD model is a reduced form of the multifluid model 
in terms of global dynamic variables describing various large 
spatial scale $(L>r_{L})$ and slow dynamical behaviors 
$(\omega<\omega_{ci})$~\cite{freidberg_2007}, where, $r_{L}$ is Larmor 
radius and $\omega_{ci}$ is ion cyclotron frequency~\cite{KIM2004448}. 
% In other words, one can ignore effect of various terms in the  
% kinetic-multifluid model which are acted on much smaller scales 
% than the MHD limit. 
% Now, we redefine a similar 
% model from the above multifluid model with more emphasis on 
% the presence of additional thermalized alpha particles 
% in addition to the electrons, ions and energetic alpha particle species.
% In present analysis, we re-define the existing Kinetic-MHD Hybrid Model by 
% including the contributions of energetic particles ($n_f,~u_f$) and 
% thermalized alpha particles ($n_0,~u_0$) to the bulk plasma electrons 
% and ions. Thus, for a general case of ($n_{\alpha}(=n_f+n_0)\sim n_b$), 
% we can have the global dynamical variables as follow,
% \begin{eqnarray}
% \rho &=& n_e m_e + n_i m_i+ n_f m_f + n_0 m_0 \\
% {\bf J} &=& q_e n_e {\bf v}_e +q_i n_i {\bf v}_i+ 
% q_f n_f {\bf u}_f + q_o n_0 {\bf u}_0\\ 
%  0 &=& q_e n_e + q_i n_i+ q_f n_f + q_0 n_0 \label{nutrality}
% \end{eqnarray}
From the continuity equation for each species given in 
Eq.~(\ref{bulk_plasma_continuity}), we multiply it by each species 
mass $m_j$ (or charge $q_j$) and summing 
over all the species, thus the mass (or charge) continuity 
equation in term of global variables ${\rho_m}$, ${\bf V}$, 
and ${\bf J}$ can be obtained as follows, 
\begin{eqnarray}
 \frac{\partial \rho_m}{\partial t} + \nabla\cdot(\rho_m{\bf V}) &=& 0,\\
 \label{single_continuity}
  and~~\frac{\partial \rho_q}{\partial t} + \nabla\cdot{\bf J} &=& 0, 
 \end{eqnarray}
where the average mass density ${\rho_m}$, the charge density ${\rho_q}$, the 
center of mass bulk velocity ${\bf V}$, 
and the total current ${\bf J}$ are defined as,  
\begin{eqnarray}
{\rho_m} = \sum_{j}^{} n_j m_j,~~ {\rho_q} = \sum_{j}^{} n_j q_j\approx 0,
\label{bulk_para} \\
{\rho_m\bf V} = \sum_{j}^{} n_j m_j {\bf v}_j,~~and~~ 
{\bf J} = \sum_{j}^{} n_j q_j {\bf v}_j. 
\end{eqnarray}
% Using the same method to 
From the momentum Eq.~(\ref{bulk_plasma_momentum}) for 
electrons and ions, we have,
\begin{eqnarray}\nonumber
{n_e m_e\frac{d {\bf v}_e}{d t}}+ n_im_i\frac{d {\bf v}_i}{d t}
= -\nabla{(p_e+p_i)} +(q_e n_e+q_i n_i){\bf E}\\\nonumber 
+ (q_en_e{\bf v}_e +q_in_i{\bf v}_i)\times{\bf B} 
+ f_{e\alpha}+ f_{i\alpha},
\label{bulk} 
\end{eqnarray}
where $f_{e\alpha}$ and $f_{i\alpha}$ are Coulomb interactions factor 
of electrons and ions with the alpha particles respectively. Now, applying 
the MHD approximations ($m_e\ll m_i$) and quasinutrality, the 
above equation can be expressed in term of global variables as follows,      
\begin{eqnarray}\nonumber
% \cancel{n_e m_e\frac{d {\bf v}_e}{d t}}+
n_im_i\frac{d {\bf v}_i}{d t}
= -\nabla{(p_e+p_i)} -(q_\alpha n_\alpha){\bf E}\\ 
+ ({\bf J}-{\bf J}_\alpha)\times{\bf B}
+ f_{e\alpha}+ f_{i\alpha}.~~~~~
\label{alpha_current_coupling} 
\end{eqnarray}
In the absence of Coulomb interactions with the $\alpha$ 
particles, (i.e., if $f_{e\alpha}$ and $f_{i\alpha}$ are neglected), 
the above Eq.~(\ref{alpha_current_coupling}) represents the 
current-coupling model discussed in 
section~\ref{Existing_model}, Eq.~(\ref{current_coup}). 
It demonstrates that 
the current-coupling model is restricted to a regime where the 
contributions of alpha particles such as the inertial and 
the Coulomb interaction effects are negligible.
% The equation does not consider the intertial effect of the 
% minor energetic $\alpha$ particles only in the total charge and 
% current density. However, the effects of 
% $\alpha$ particle's inertia 
% are not considered in the equation. 
% This means, the current-coupling scheme is not sufficient for 
% general cases having substantial amount of alpha particles.
% In the limit of ($n_{\alpha}\sim n_e ~(or~n_i)$), the equilibration 
% force 
% balance will be modified including the eq alpha particles 
% distribution and interacting forces.

Now, in presence of significant amount of the alpha particles both 
thermalized helium ash and energetic particles with density 
($n_{\alpha}=n_0+n_f)$ comparable to the background plasma density ($n_b$), 
the sum of momentum equation for both the species becomes, 
 \begin{eqnarray}\nonumber
 m_f\frac{\partial n_f{\bf v}_f}{\partial t}+n_0m_0\frac{d {\bf v}_0}{d t}
% n_fm_f\left(\frac{\partial {\bf u}_f}{\partial t}+{\bf u}_f\cdot\nabla{\bf u}_f\right) 
= -\nabla\cdot{\bf P_f}-\nabla p_0 +(q_f n_f+q_0 n_0){\bf E}\\ \nonumber
+ (q_fn_f{\bf v}_f +q_0n_0{\bf v}_0)\times{\bf B}
+ f_{\alpha e}+ f_{\alpha i}. ~~~~~~~~~
\label{alpha_coup} 
\end{eqnarray}
Where, ${\bf v}_f$ and ${\bf v}_0$ are the average velocities of 
energetic particles and thermalized helium ash respectively. 
Applying the general conditions, 
$n_f\ll n_0$, $m_0=m_f (\approx m_\alpha)$, and 
$q_0=q_f (\approx q_\alpha)$, the equation can be rewritten 
as follows,  
\begin{eqnarray}\nonumber
% \cancel{m_f\frac{\partial n_f{\bf v}_f}{\partial t}}
{n_0m_0\frac{d {\bf v}_0}{d t}}
% n_fm_f\left(\frac{\partial {\bf u}_f}{\partial t}+{\bf u}_f\cdot\nabla{\bf u}_f\right) 
= -\nabla\cdot{\bf P_f}-\nabla p_0  
+(q_\alpha n_\alpha){\bf E}\\ + {\bf J}_\alpha\times{\bf B}
+ f_{\alpha e}+ f_{\alpha i}. ~~~~~~~~~
\label{alpha_coup} 
\end{eqnarray}
Further, coupling the above two equations i.e., 
Eq.~(\ref{alpha_current_coupling}) for the background plasma and 
Eq.~(\ref{alpha_coup}) for the alpha particles into a 
single equation, and applying ($f_{ij}=-f_{ji})$, the total momentum 
balance equation becomes, 
\begin{eqnarray}\nonumber
 n_im_i\frac{d{\bf v}_i}{d t}+n_0m_0\frac{d{\bf v}_0}{d t}
= -\nabla\cdot{\bf P_f}-\nabla p + {\bf J}\times{\bf B}. 
\label{alpha_bulk_coup} 
\end{eqnarray}
Where,  $p=p_e +p_i+p_0$ is the sum of the separate partial pressure of 
all the thermalized species. 
% Now, the major difficulty is the coupling of the 
% inertial non-linear terms in the left hand side which does not 
% lend itself into the global variables. However, 
Using the MHD limits and general approximations 
that $m_e\ll m_i$, $n_0<n_i$, and 
${\bf v}_0\approx {\bf v}_i$ in the 
relation ${\rho_m\bf V} = \sum_{j}^{} n_j m_j {\bf v}_j$, the above 
equation can be reduced as follows, 
\begin{eqnarray}
 \rho_m\frac{d{\bf V}}{d t}
= -\nabla\cdot{\bf P_f}-\nabla p + {\bf J}\times{\bf B}.
\label{alpha_pressure_coupling} 
\end{eqnarray}
This new equation becomes a modified 
form of the pressure-coupling scheme~\cite{doi:10.1063/1.860011,KIM2004448}, 
with contributions of alpha particles in all the global dynamical 
variables $\rho_m$, ${\bf V}$, $p$, and ${\bf J}$. It further demonstrated 
that the pressure-coupling model Eq.~(\ref{alpha_pressure_coupling}) 
is more extensive than the current-coupling model recovered in 
Eq.~(\ref{alpha_current_coupling}), and useful for 
cases having a substantial amount of thermalized helium ash and 
energetic alpha particles. 
% 
% Further, this equation shows that a steady state 
% equilibrium ($\rho_m\frac{d{\bf V}}{d t}=0$) is
% achieved when the magnetic force balances the total pressure 
% force even in the presence of thermalized and energetic 
% alpha particles.

Now, for the energy equation, if we ignore the various sources in the
multifluid energy equation~(\ref{bulk_plasma_energy}) 
% which are acted on a much smaller time scale than 
using the MHD limit. The energy equation in terms of 
global variables can be written as follows, 
\begin{eqnarray}\nonumber
\frac{1}{\gamma-1}\left(\frac{dp}{dt}\right) =-p\nabla\cdot{\bf V}
  -\nabla\cdot {\bf q}, \\
%  \pm S_j \pm\sum_{k}^{} 2 n_jm_j\frac{m_j}{m_k}\nu_{jk}({T_j}-{T_k}),  
\label{MHD_plasma_energy} 
\end{eqnarray}
where, $\gamma$ is sum of $\gamma_j$, $p$ is the total pressure, 
and ${\bf q}$ is the heat flux due to thermal 
conduction~\cite{doi:10.1063/1.873437}.
% This single fluid equations suffers for being a non-Hamiltonian in 
% nature and unable to conserve total energy~\cite{Tronci_2014}. 
% It may be due to continuous loss of energetic particles from the 
% burning plasma system~\cite{doi:10.1063/1.4890955}. However, the 
% assumption of relatively less energetic particles $n_f\ll n_0$ would 
% nearly conserved the energy. 
Furthermore, the associated Maxwell's equations in the MHD 
limit are as follows, 
\begin{eqnarray}
\frac{\partial {\bf B}}{\partial t} &=& -\nabla\times {\bf E} ,\\
{\bf J} &=& \frac{1}{\mu_0}\nabla\times{\bf B} ,\\
\nabla\cdot{\bf B} &=& 0.
\end{eqnarray}

The other important relation is the generalized Ohm's law that relates 
the electric field ${\bf E}$ to the global fluid variables ${\bf V}$ 
and ${\bf J}$. It is found that the Ohm's law is modified in the presence 
of the substantial amount of alpha particles in the system.
The full derivation of the extended generalized Ohm's law for 
such a multifluid system is given in Appendix~\ref{appendixA}, and 
the final equation is given here, 
\begin{widetext}
\begin{eqnarray} \nonumber
({\bf E}+ {\bf V}\times {\bf B}) =
\left(\frac{n_{\alpha} m_{\alpha}}{\rho_m}
-n_{\alpha}q_{\alpha}\frac{m_im_e}{q_iq_e\rho_m}(\frac{q_i}{m_i} 
+\frac{q_e}{m_e}-\frac{q_{\alpha}}{m_{\alpha}})\right)
({\bf E}+ {\bf v}_{\alpha}\times {\bf B})
-\frac{m_im_e}{q_iq_e\rho_m}\frac{\partial {\bf J}}{\partial t} \\ \nonumber
 +\frac{m_im_e}{q_iq_e\rho_m}(\frac{q_e}{m_e}+\frac{q_i}{m_i})
 ({\bf J}\times {\bf B}) 
 -(\frac{m_im_e}{q_iq_e\rho_m})\bigg(\frac{q_i}{m_i}\nabla{p_i}+
 \frac{q_e}{m_e}\nabla{p_e}
 +\frac{q_{\alpha}}{m_{\alpha}}\nabla{p_{\alpha}}\bigg) \\ \nonumber
 -(\frac{m_im_e}{q_iq_e\rho_m} )\bigg[{\bf J}(\nu_{i{\alpha}}
 +\nu_{{\alpha}e}+\nu_{ei}) +\rho_m{\bf V}(\frac{q_i}{m_i}\nu_{ie} 
 +\frac{q_e}{m_e}\nu_{e{\alpha}}
 +\frac{q_{\alpha}}{m_{\alpha}}\nu_{{\alpha}i})\\ \nonumber
 - \nu_{{\alpha}i}\bigg((n_{\alpha}q_{\alpha}
 +\frac{n_im_iq_{\alpha}}{m_{\alpha}}){\bf v}_i 
 + (\frac{n_{\alpha}m_{\alpha}}{n_im_i}n_eq_e 
 +\frac{n_em_eq_{\alpha}}{m_{\alpha}}){\bf v}_e 
 -(\frac{n_{\alpha}m_{\alpha}}{n_im_i}n_eq_e){\bf v}_{\alpha}\bigg) \\ \nonumber
 - \nu_{ie}\bigg((-\frac{n_im_i}{n_em_e}n_{\alpha}q_{\alpha}){\bf v}_i 
  +(n_iq_i+\frac{n_em_eq_i}{m_i}) {\bf v}_e 
 +(\frac{n_im_i}{n_em_e}n_{\alpha}q_{\alpha} 
 +\frac{n_{\alpha}m_{\alpha}q_i}{m_i}){\bf v}_{\alpha}\bigg)\\
 -\nu_{e{\alpha}}\bigg((\frac{n_em_e}{n_{\alpha}m_{\alpha}}n_iq_i  
 +\frac{n_im_iq_e}{m_e}){\bf v}_i 
 -(\frac{n_em_e}{n_{\alpha}m_{\alpha}}n_iq_i){\bf v}_e+(n_eq_e
 +\frac{n_{\alpha}m_{\alpha} q_e}{m_{\alpha}}){\bf v}_{\alpha}\bigg)\bigg].~~~
\label{generalised_Ohms_law}
\end{eqnarray}
\end{widetext}
%%%%%%%%%%%%%%%%%%%%%%%%%
%
The above equation has shown how the Ohm's law is modified due to 
the presence of $\alpha$ particles in the burning plasma.
This equation relates the electric field to the density and 
fluid velocity of all multi-species present in the system.
Specifically, the left-hand side represents the electric field 
in the reference frame moving with the plasma. 
Again, in the right-hand side, the first term arises mainly due to 
the presence of the new component alpha species, the second term arises 
from the inertial flows, the third term arises because of Hall current 
across the magnetic field, the fourth term is due to pressure 
variations of all the species, and the last term is due to Coulomb 
collisions which give the resistivity effects in the multi-species 
flow system. In the general MHD approximations that $m_e\ll m_i$, 
$\frac{q_e}{m_e}\gg \frac{q_i}{m_i}$, and ${p_e}\sim{p}_i$, 
the order of the inertial term is relatively small and negligible. 
The Hall term and pressure term are of the same order, but 
both  are negligible for low frequencies process. However, 
inertial terms and Hall term are very important if one wants 
to properly capture the plasma dynamics in tokamak MHD 
modeling~\cite{gourdain2017impact}. 
The resistivity terms which represent the effects of the substantial 
amount of alpha particles are important in such a 
multi-species system~\cite{doi:10.1029/1998JA900065}.

The above equation for extended generalized Ohm's law 
readily reduces to the following equation for two-fluid 
plasma if the contributions from the alpha particles are ignored. 
% For example, when we apply the condition $n_\alpha$, $q_\alpha$, and 
% ${\bf v_\alpha}$ goes to zero in the above equation.
% Then, the equation becomes, 
 \begin{widetext}
\begin{eqnarray} \nonumber
({\bf E}+ {\bf V}\times {\bf B}) =
-\frac{m_im_e}{q_iq_e\rho_m}\frac{\partial {\bf J}}{\partial t} 
 +\frac{m_im_e}{q_iq_e\rho_m}(\frac{q_e}{m_e}+\frac{q_i}{m_i})
 ({\bf J}\times {\bf B}) 
 -(\frac{m_im_e}{q_iq_e\rho_m})\bigg(\frac{q_i}{m_i}\nabla{p_i}+
 \frac{q_e}{m_e}\nabla{p_e}\bigg) \\ \nonumber
 -(\frac{m_im_e}{q_iq_e\rho_m} )\bigg[{\bf J}\nu_{ei} 
 +\rho_m{\bf V}(\frac{q_i}{m_i})\nu_{ie}
 - \nu_{ie}\bigg(n_iq_i+\frac{n_em_eq_i}{m_i}\bigg) {\bf v}_e\bigg],
\label{generalised_Ohms_law}
\end{eqnarray}
% \end{widetext}
\begin{eqnarray} \nonumber
\Rightarrow ({\bf E}+ {\bf V}\times {\bf B}) =
-\frac{m_im_e}{q_iq_e\rho_m}\frac{\partial {\bf J}}{\partial t}
+\frac{m_im_e}{q_iq_e\rho_m}(\frac{q_e}{m_e}+\frac{q_i}{m_i})
({\bf J}\times {\bf B})  \\ \nonumber
-(\frac{m_im_e}{q_iq_e\rho_m})\bigg(\frac{q_i}{m_i}\nabla{p_i}+
\frac{q_e}{m_e}\nabla{p_e}\bigg) 
-\frac{m_im_e}{q_iq_e\rho_m}
\bigg(1+\frac{n_e m_e}{n_i m_i} \bigg) \nu_{ie}{\bf J}.
\end{eqnarray}
 \end{widetext}
This is an expanded form of the generalized Ohm's law for the 
two-fluid plasma~\cite{doi:10.1029/1998JA900065,
doi:10.1002/ctpp.19650050103}. Further, dropping the inertia, Hall term, and 
pressure terms in the resistive MHD limit, the equation reduces to the 
simplified Ohm's law as follows,
\begin{align} \nonumber
({\bf E}+ {\bf V}\times {\bf B}) \approx \frac{m_im_e}{q_iq_e\rho_m}
\bigg(\frac{n_i m_i+n_e m_e}{n_i m_i} \bigg) \nu_{ie}{\bf J},\\ \nonumber
 \approx \frac{m_im_e}{q_iq_e}
\bigg(\frac{1}{n_i m_i} \bigg) \nu_{ie}{\bf J},~~~~~~~~~~~\\ \nonumber
 \approx \eta {\bf J},~~~~~~~~~~~~~~~~~~~~~~~~~~~~~~~~
\end{align}
% Using value of $\rho_m$ from the Eq.~(\ref{bulk_para}), we get,
% \begin{eqnarray} 
% ({\bf E}+ {\bf V}\times {\bf B}) \approx -\frac{m_im_e}{q_iq_e}
% \bigg(\frac{1}{n_i m_i} \bigg) \nu_{ie}{\bf J},\\ \nonumber
% \approx \eta {\bf J},~~~~~~~~~~~~~~~~~~~~~~
% \end{eqnarray}
where, $\eta =\frac{m_e\nu_{ie}}{n_i q_e q_i }$ is the resistivity of 
two-fluid plasma. Now, it is to be noted that, the above single-fluid 
equations~(\ref{alpha_pressure_coupling}) to (\ref{generalised_Ohms_law}) 
are derived for the general ignited burning plasma that 
has contributions from both non-thermalized energetic alpha 
particles and a major amount of thermalized helium 
ash, i.e., $n_{\alpha}= (n_0 +n_f) \approx n_b$ and $n_0\gg n_f$. 
These equations can be used to check whether the alpha pressure gradient 
will excite instabilities that would cause the alphas particles lost at 
a fast rate. Besides, a significant amount of helium ash will 
dilute the D–T fuel, leading to a reduction in fusion reactions. 

Further, 
when the contribution of the thermalized particles is reduced, i.e., 
$n_0\approx 0$ and $n_f\ll n_b$, the whole single-fluid model tends 
to a Non-Hamiltonian form of the pressure coupling model 
(derived by C .Tronci {\it~et~al}~\cite{Tronci_2014}). 
The non-Hamiltonian nature is due to the continuous loss of energetic 
particles from the plasma system. But, the density condition of relatively 
less energetic particles $n_f\ll n_0$ would lead to a nearly 
conservation of the total energy. Under the same density regime, 
the model is completely equivalent to the current coupling model as 
discussed in above section~\ref{Existing_model}.
% 
% where 
% $n_{\alpha}= (n_0 +n_f) \approx n_b$, with all the possible regimes 
% $n_0\ll n_f$, $n_0\approx n_f$, or $n_0\gg n_f$.
% 
%  
% % many multi-species 
% % system in space plasma at different levels of chosen time and 
% % length scales~\cite{SONG2005447}. 
%  These equations can be used for studies of global
% low-frequency behaviors in multi-fluid systems such as 
% burning plasma and space plasmas. 
% % But, these issues can be ignored under the approximations that 
% % the amount of loss energetic alpha particles is relatively small. 
% % Furthermore, the equations can be closed either by defining the 
% % unknown variables $q$ or ignored it in the MHD 
% % limit~\cite{doi:10.1063/1.873437}.
% % as the heating and 
% % cooling of the whole plasma take place on a much slower time 
% % scales 
% 
% 
% When real geometric effects are important the two-fluid equations are
% often too complicated to solve.
% Various forms of the MHD model are useful for the study of 

% %%%%%%%%%%%%%%%%%%%%%%%%%%%%%%%%%%%%%%%%%%%%%%%%%%%%%%%%%%%%%%%%%%%
\section{Summary and Conclusions}
\label{conclusion} 
We start from a brief review of the pressure and current-coupling model 
of kinematic-MHD hybrid models for alpha particle physics in burning plasma. 
These equivalent models are limited to specific parameter regimes where 
energetic alpha particles are rarefied and their pressure is comparable 
to that of the bulk plasma.
% i.e., $n_{\alpha}\ll n_b$ 
% and $\beta_{\alpha}\approx \beta_b$. 
Besides, there is no 
consideration of thermalized helium ash in the existing models.
% In addition, 
% these models are deficient regarding the use of adiabatic 
% energy equation and ideal Ohm's las even in the presence of alpha particles. 
We extend the existing model to consider the contributions from both the 
non-thermalized alpha particles and a major amount of 
thermalized helium ash present in the ignited burning 
plasma, where $n_{\alpha}= (n_0 +n_f) \approx n_b$, with all the possible 
regimes $n_0\ll n_f$, $n_0\approx n_f$, or $n_0\gg n_f$. 
In this kinetic-multifluid 
% model, for account of all the possible dynamical regime of the system. 
model, the non-thermalized particles are treated by the kinetic approach, 
and the thermalized particles are described as fluid. 
% In particular, 
% we approximate a new variable, for the non-thermalized particles, which 
% reduces to the role of temperature in the energy equation. T
This model describes explicitly the role of each dynamical variables 
and system parameters in a wide range of varying length scale 
and time scale. 

% that has demonstrated various 
% limitations against the equivalent of the Pressure-coupling model to the 
% Current-coupling model.
%
The kinetic-multifluid model is further reduced to an 
extended hybrid kinetic-MHD model in the MHD limit. 
This reduction process demonstrates 
that the existing pressure-coupling model is more extensive than 
the current-coupling scheme and may be sufficient for the regime 
% having both 
% the energetic and enormous amount of helium ash with density 
where $n_0\gg n_f$, 
whereas the kinetic-multifluid model is still required in the regime 
$n_0\approx n_f$. The single-fluid MHD limit of the model is derived 
to account the global impact of the substantial amount of alpha 
particles in the low-frequency regime of burning plasma. In the model, 
substantial changes are introduced to the Ohm's law due to the presence 
of the major amount of alpha particles in the system. Further, when the 
contribution of the thermalized particles is 
neglected, i.e., $n_0\approx 0$ and $n_f\ll n_b$, the model reduces 
to a Non-Hamiltonian form of the pressure coupling model which is 
again equivalent to the existing current coupling model.
% useful for describing the impact of alpha particles to low-frequency 
% global behavior of the burning plasma.
In the future, we plan on implementing these extended hybrid models 
in Non-Ideal MHD with Rotation (NIMROD) code 
and studies the impact of the alpha particles presence in tokamak system. 
% In conclusion, we will implement these hybrid models in 
% Non-Ideal MHD with Rotation (NIMROD) code 
% and studies the impact of the alpha particles presence in tokamak system. 
% % It could be used to 
% re-examine low-frequency global behaviors and the nature of instabilities driven 
% by the energetic particles in the ignited burning plasma. 
Besides fusion plasma, 
the proposed models can be used for studies of low-frequency processes 
in space 
plasma such as the interaction of energetic solar wind with 
the magnetosphere, the ionosphere, and the thermosphere. 
% These formulations may further extend up to a 
% wide range of parameters regime where 
% the real experiments system can not even reach. 
%%%%%%%%%%%%%%%%%%%%%%%%%%%%%%%%
\section{Acknowledgments}
Authors acknowledge the support of the State Administration of
Foreign Experts Affairs­ - Foreign Talented Youth Introduction Plan
under Grant No. WQ2017ZGKX065. 
% This work was supported by the Fundamental Research Funds for 
% the Central Universities at Huazhong University of Science and 
% Technology Grant No. 2019kfyXJJS193, the National Natural Science 
% Foundation of China Grant No. 11775221, the U.S. DOE 
% Grant Nos. DE-FG02-86ER53218 and DE-SC0018001.
% 
% This research was supported by the
% National Natural Science Foundation of China Grant No. 11775221.
% Author P. Zhu also acknowledges the supports from U.S. DOE 
% Grant Nos. DE-FG02-86ER53218 and DE-SC0018001. 
% The numerical calculations in this paper used
% resources of Supercomputing Center of University of Science and
% Technology of China. 
%%%%%%%%%%%%%%%%%%%%%%%%%%%%%%%
\begin{widetext}

\appendix
\section{Derivation of the generalized Ohm's law
\label{appendixA}}
% \cite{doi:10.1029/1998JA900065}
% The full derivation of the modified Ohm's law is given below,

Starting from the momentum equation for each $j-$species in the 
multi-fluid model given in Eq.~(\ref{bulk_plasma_momentum}) as follows, 
\begin{eqnarray}
n_im_i\frac{d {\bf v}_i}{d t}
= \pm \eta_i n_im_i{\bf v}_i  -\nabla{p_i} 
+q_i n_i({\bf E}+ {\bf v_i}\times{\bf B})
\pm\sum_{j}^{} m_i n_i\nu_{ij}({\bf v_i}-{\bf v_j})~~~~
% \pm m_i n_i\nu_{ik}({\bf v_i}-{\bf v_k}) 
% \pm m_i n_i\nu_{if}({\bf v_i}-{\bf v_f}) 
\label{bulk_2} 
\end{eqnarray}
Then, we also redefine the global dynamical variables of the 
multi-species system as,
\begin{eqnarray}
\rho_m &=& n_i m_i + n_j m_j+ n_k m_k \\
\rho_m {\bf V} &=& n_im_i{\bf v}_i  + 
n_jm_j{\bf v}_j + n_k m_k{\bf v}_k  \\
{\bf J} &=& q_i n_i {\bf v}_i +q_j n_j {\bf v}_j+ 
q_k n_k {\bf u}_k \\ 
 0 &=& q_i n_i + q_j n_j+ q_k n_k 
\label{nutrality_2}
\end{eqnarray}

Here, $i$, $j$, and $k$ represent electrons, ions, and alpha particles. 
$n_k(=n_f+n_0)$ includes the contribution from both 
energetic particles ($n_f$) and thermalized alpha particles ($n_0$). 
The generation or annihilation term ($\pm \eta_j n_jm_j{\bf v}_j$) 
is ignored in the MHD limit. Then, we multiply the 
equation~(\ref{bulk_2}) by $\frac{q_i}{m_i}$ and 
take sum over all the $i-$ species. We get, 
\begin{eqnarray} \nonumber
\sum_{i}^{}
n_iq_i\frac{d {\bf v}_i}{d t}
= \sum_{i}^{}\left[
-\frac{q_i}{m_i}\nabla{p_i} +\frac{{q_i}^2 n_i}{m_i}({\bf E}+ {\bf v_i}\times{\bf B})
% \right. \\ \left.
\pm \frac{q_i}{m_i}\sum_{j}^{} m_i n_i\nu_{ij}
({\bf v_i}-{\bf v_j})\right].
% \pm m_i n_i\nu_{ik}({\bf v_i}-{\bf v_k}) 
% \pm m_i n_i\nu_{if}({\bf v_i}-{\bf v_f}) 
\label{ohms_law_1} 
\end{eqnarray}
% Where, $i$ is the no. of species present in the system.
The difficulty with the left hand side (LHS) is that the sum over of 
nonlinear convective terms is not easily lend itself into a single 
global variables. LHS can be express as 
$[\frac{\partial {\bf J}}{\partial t} +\nabla\cdot ({\bf VJ+JV})]$ in 
more complete form~\cite{KIM2004448}. However, using linearize approximation, 
we ignore the convective part and thus LHS is written as 
$\frac{\partial {\bf J}}{\partial t}$ for the present analysis. 
For the right hand side (RHS), we use quasi-neutrality condition 
given in Eqn~(\ref{nutrality_2}), to obtain,
\begin{eqnarray}\nonumber
 1^{st}~term &=&  \sum_{i}^{}-\frac{q_i}{m_i}\nabla{p_i} \\ 
 &=& -\bigg(\frac{q_i}{m_i}\nabla{p_i}+\frac{q_j}{m_j}\nabla{p_j}
 +\frac{q_k}{m_k}\nabla{p_k}\bigg).
\end{eqnarray}

\begin{eqnarray}\nonumber
 2^{st}~term &=& \sum_{i}^{}\frac{{q_i}^2 n_i}{m_i}{\bf E}\\ \nonumber
%   &=& (\frac{{q_i}^2 n_i}{m_i}+\frac{{q_j}^2 n_j}{m_j}
%   +\frac{{q_k}^2 n_k}{m_k} ){\bf E} \\ \nonumber
  &=& (n_iq_i\frac{q_i}{m_i}+n_jq_j\frac{q_j}{m_j}
  +n_kq_k\frac{q_k}{m_k}){\bf E} \\ \nonumber
  &=& \left( (-n_jq_j-n_kq_k)\frac{q_i}{m_i}
      +(-n_iq_i-n_kq_k)\frac{q_j}{m_j}
      +n_kq_k\frac{q_k}{m_k} \right){\bf E}  \\ \nonumber
  &=& \left(-\frac{n_jq_jm_jq_i+n_iq_im_iq_j}{m_im_j} 
               -\frac{n_kq_km_jq_i+n_kq_km_iq_j}{m_im_j} 
       +n_kq_k\frac{q_k}{m_k} \right){\bf E}  \\ 
         &=& \left( -\frac{q_jq_j}{m_im_j}\rho_m 
               +\frac{q_jq_j}{m_im_j}{n_km_k} 
       -n_kq_k(\frac{q_i}{m_i} +\frac{q_j}{m_j}-\frac{q_k}{m_k})
       \right){\bf E}.
\end{eqnarray}

\begin{eqnarray}\nonumber
 3^{th}~term &=& \sum_{i}^{}\frac{{q_i}^2 n_i}{m_i}{\bf v}_i\times {\bf B}\\ \nonumber
%   &=& (\frac{{q_i}^2 n_i}{m_i}+\frac{{q_j}^2 n_j}{m_j}
%   +\frac{{q_k}^2 n_k}{m_k} ){\bf E} \\ \nonumber
  &=& \bigg(\frac{n_i{q_i}^2{\bf v}_i}{m_i}+\frac{n_j{q_j}^2{\bf v}_j}{m_j}
  +\frac{n_k{q_k}^2{\bf v}_k}{m_k}\bigg)\times {\bf B} \\ \nonumber
  &=& \bigg[\frac{q_iq_j}{m_im_j} \left( {n_iq_i{\bf v}_i}\frac{m_j}{q_j} 
  +{n_jq_j{\bf v}_j}\frac{m_i}{q_i} \right) 
  +\frac{n_k{q_k}^2{\bf v}_k}{m_k} \bigg]\times {\bf B}\\ \nonumber
  &=& \bigg[-\frac{q_iq_j}{m_im_j} 
  \left( ({n_im_i{\bf v}_i}+ {n_jm_j{\bf v}_j})
  -(\frac{m_j}{q_j}+\frac{m_i}{q_i}){n_iq_i{\bf v}_i} 
  -(\frac{m_i}{q_i}+\frac{m_j}{q_j}){n_jq_j{\bf v}_j} \right) 
  +\frac{n_k{q_k}^2{\bf v}_k}{m_k} \bigg]\times {\bf B}\\ \nonumber
% %  &=& \bigg[-\frac{q_iq_j}{m_im_j}(\rho_m{\bf V}-n_km_k{\bf v}_k)
% %   +(\frac{q_j}{m_j}+\frac{q_i}{m_i})({\bf J}-n_kq_k{\bf v}_k)
% %   +\frac{n_k{q_k}^2{\bf v}_k}{m_k} 
% %   \bigg]\times {\bf B}\\ 
   &=& \bigg[-\frac{q_iq_j}{m_im_j}\rho_m{\bf V} 
   +\frac{q_iq_j}{q_iq_j}n_km_k{\bf v}_k
  +(\frac{q_j}{m_j}+\frac{q_i}{m_i}){\bf J}
  -n_kq_k{\bf v}_k(\frac{q_i}{m_i}+\frac{q_j}{m_j}-\frac{q_k}{m_k}) 
  \bigg]\times {\bf B}.
\end{eqnarray}

and,

\begin{eqnarray}\nonumber
 4^{th}~term &=& \sum_{i}^{} \frac{q_i}{m_i}
\sum_{j}^{} m_i n_i\nu_{ij}({\bf v_i}-{\bf v_j}) \\ \nonumber
&=& \sum_{i}^{} \bigg( n_i q_i\nu_{ik}({\bf v_i}-{\bf v_k})
-n_i q_i\nu_{ij}({\bf v}_i-{\bf v_j}) \bigg) \\ \nonumber
&=& \sum_{i}^{} \bigg( -\nu_{ik}({\bf J}- n_jq_j{\bf v}_j
-n_kq_k{\bf v}_k-n_iq_i{\bf v}_k) 
+\nu_{ij}(n_iq_i{\bf v_j}-n_iq_i{\bf v_i}) \bigg) 
\end{eqnarray}
Further, summing over all the $i$-species, we get, 
\begin{eqnarray}\nonumber
  4^{th}~term = -{\bf J}(\nu_{ik}+\nu_{kj}+\nu_{ji})
+ \nu_{ik}(n_jq_j{\bf v}_j +n_kq_k{\bf v}_k+n_iq_i{\bf v}_k)
+ \nu_{ij}(n_iq_i{\bf v}_j -n_iq_i{\bf v}_i)   \\ \nonumber
 + \nu_{ji}(n_iq_i{\bf v}_i +n_kq_k{\bf v}_k+n_iq_i{\bf v}_i)
 + \nu_{jk}(n_jq_j{\bf v}_k -n_jq_j{\bf v}_j) 
 + \nu_{kj}(n_iq_i{\bf v}_i +n_jq_j{\bf v}_j+n_kq_k{\bf v}_j)  \\ \nonumber
 + \nu_{ki}(n_kq_k{\bf v}_i -n_kq_k{\bf v}_k)  
\end{eqnarray}
\begin{eqnarray}\nonumber
= -{\bf J}(\nu_{ik}+\nu_{kj}+\nu_{ji})
+ \nu_{ik}(n_jq_j{\bf v}_j +n_kq_k{\bf v}_k+n_iq_i{\bf v}_k) 
+ \nu_{ij}\frac{q_i}{m_i}(n_im_i{\bf v}_j 
-\rho_m{\bf V}+ n_jm_j{\bf v}_j+n_km_k{\bf v}_k)  \\ \nonumber
 + \nu_{ji}(n_iq_i{\bf v}_i +n_kq_k{\bf v}_k+n_iq_i{\bf v}_i)
 + \nu_{jk}\frac{q_j}{m_j}(n_jm_j{\bf v}_k 
 -\rho_m{\bf V}+ n_im_i{\bf v}_i+n_km_k{\bf v}_k) \\ \nonumber
 + \nu_{kj}(n_iq_i{\bf v}_i +n_jq_j{\bf v}_j+n_kq_k{\bf v}_j) 
 +\nu_{ki}\frac{q_k}{m_k}(n_km_k{\bf v}_k 
 -\rho_m{\bf V}+ n_im_i{\bf v}_i+n_jm_j{\bf v}_j) 
 \end{eqnarray}
 \begin{eqnarray}\nonumber
= -{\bf J}(\nu_{ik}+\nu_{kj}+\nu_{ji}) 
-\rho_m{\bf V}(\frac{q_i}{m_i}\nu_{ij} +\frac{q_j}{m_j}\nu_{jk}
+\frac{q_k}{m_k}\nu_{ki} ) 
+ \nu_{ki}\bigg((n_kq_k+\frac{n_im_iq_k}{m_k}){\bf v}_i 
+ (\frac{n_km_k}{n_im_i}n_jq_j  \\ \nonumber 
+\frac{n_jm_jq_k}{m_k}){\bf v}_j 
 -(\frac{n_km_k}{n_im_i}n_jq_j){\bf v}_k\bigg)  
 + \nu_{ij}\bigg((-\frac{n_im_i}{n_jm_j}n_kq_k){\bf v}_i 
  +(n_iq_i+\frac{n_jm_jq_i}{m_i}) {\bf v}_j
 +(\frac{n_im_i}{n_jm_j}n_kq_k  \\ \nonumber
 +\frac{n_km_kq_i}{m_i}){\bf v}_k\bigg)
 +\nu_{jk}\bigg((\frac{n_jm_j}{n_km_k}n_iq_i +\frac{n_im_iq_j}{m_j}){\bf v}_i
 -(\frac{n_jm_j}{n_km_k}n_iq_i){\bf v}_j
 +(n_jq_j+\frac{n_km_k q_j}{m_k}){\bf v}_k\bigg)
\end{eqnarray}

Now, regrouping all the terms in Eqn~(\ref{ohms_law_1}), we can write, 
\begin{eqnarray} \nonumber
\frac{\partial {\bf J}}{\partial t}
= -\bigg(\frac{q_i}{m_i}\nabla{p_i}+\frac{q_j}{m_j}\nabla{p_j}
+\frac{q_k}{m_k}\nabla{p_k}\bigg)
+\left( -\frac{q_iq_j}{m_im_j}\rho_m 
               +\frac{q_iq_j}{m_im_j}{n_km_k} 
       -n_kq_k(\frac{q_i}{m_i} +\frac{q_j}{m_j}-\frac{q_k}{m_k})
       \right){\bf E} \\ \nonumber
 +\bigg(-\frac{q_iq_j}{m_im_j}\rho_m{\bf V} 
    +\frac{q_iq_j}{m_im_j}n_km_k{\bf v}_k
   +(\frac{q_j}{m_j}+\frac{q_i}{m_i}){\bf J}
   -n_kq_k{\bf v}_k(\frac{q_i}{m_i}+\frac{q_j}{m_j}-\frac{q_k}{m_k}) 
   \bigg)\times {\bf B} \\ \nonumber
 -{\bf J}(\nu_{ik}+\nu_{kj}+\nu_{ji}) 
 -\rho_m{\bf V}(\frac{q_i}{m_i}\nu_{ij} +\frac{q_j}{m_j}\nu_{jk}
 +\frac{q_k}{m_k}\nu_{ki} ) 
 + \nu_{ki}\bigg((n_kq_k+\frac{n_im_iq_k}{m_k}){\bf v}_i 
 + (\frac{n_km_k}{n_im_i}n_jq_j \\ \nonumber
 +\frac{n_jm_jq_k}{m_k}){\bf v}_j  
  -(\frac{n_km_k}{n_im_i}n_jq_j){\bf v}_k\bigg)  
 + \nu_{ij}\bigg((-\frac{n_im_i}{n_jm_j}n_kq_k){\bf v}_i 
  +(n_iq_i+\frac{n_jm_jq_i}{m_i}) {\bf v}_j
+(\frac{n_im_i}{n_jm_j}n_kq_k \\ \nonumber
+\frac{n_km_kq_i}{m_i}){\bf v}_k\bigg)
 +\nu_{jk}\bigg((\frac{n_jm_j}{n_km_k}n_iq_i +\frac{n_im_iq_j}{m_j}){\bf v}_i
 -(\frac{n_jm_j}{n_km_k}n_iq_i){\bf v}_j
 +(n_jq_j+\frac{n_km_k q_j}{m_k}){\bf v}_k\bigg).
\end{eqnarray}

\begin{eqnarray} \nonumber
\Longrightarrow\frac{m_im_j}{q_iq_j\rho_m}  \frac{\partial {\bf J}}{\partial t}
= -(\frac{m_im_j}{q_iq_j\rho_m})\bigg(\frac{q_i}{m_i}\nabla{p_i}+
\frac{q_j}{m_j}\nabla{p_j}+\frac{q_k}{m_k}\nabla{p_k}\bigg)
+\bigg(  -1 +\frac{n_km_k}{\rho_m} 
       -n_kq_k\frac{m_im_j}{q_iq_j\rho_m}(\frac{q_i}{m_i}  \\ \nonumber
       +\frac{q_j}{m_j}-\frac{q_k}{m_k})
       \bigg){\bf E} 
 +\bigg[-{\bf V}+\frac{n_km_k}{\rho_m}{\bf v}_k
   +(\frac{m_im_j}{q_iq_j\rho_m} )(\frac{q_j}{m_j}+\frac{q_i}{m_i}){\bf J}
   -n_kq_k{\bf v}_k(\frac{m_im_j}{q_iq_j\rho_m} )(\frac{q_i}{m_i}+\frac{q_j}{m_j}\\ \nonumber
   -\frac{q_k}{m_k})\bigg]\times {\bf B} 
 -(\frac{m_im_j}{q_iq_j\rho_m} )\bigg[{\bf J}(\nu_{ik}+\nu_{kj}+\nu_{ji}) 
 +\rho_m{\bf V}(\frac{q_i}{m_i}\nu_{ij} +\frac{q_j}{m_j}\nu_{jk}
  +\frac{q_k}{m_k}\nu_{ki} ) 
 - \nu_{ki}\bigg((n_kq_k  \\ \nonumber
  +\frac{n_im_iq_k}{m_k}){\bf v}_i 
 + (\frac{n_km_k}{n_im_i}n_jq_j +\frac{n_jm_jq_k}{m_k}){\bf v}_j 
 -(\frac{n_km_k}{n_im_i}n_jq_j){\bf v}_k\bigg)  
 - \nu_{ij}\bigg((-\frac{n_im_i}{n_jm_j}n_kq_k){\bf v}_i \\ \nonumber
  +(n_iq_i+\frac{n_jm_jq_i}{m_i}) {\bf v}_j
 +(\frac{n_im_i}{n_jm_j}n_kq_k +\frac{n_km_kq_i}{m_i}){\bf v}_k\bigg)
 -\nu_{jk}\bigg((\frac{n_jm_j}{n_km_k}n_iq_i +\frac{n_im_iq_j}{m_j}){\bf v}_i \\ \nonumber
 -(\frac{n_jm_j}{n_km_k}n_iq_i){\bf v}_j
  +(n_jq_j+\frac{n_km_k q_j}{m_k}){\bf v}_k\bigg) \bigg].
\end{eqnarray}

\begin{eqnarray} \nonumber
\Longrightarrow({\bf E}+ {\bf V}\times {\bf B})=
\left(\frac{n_km_k}{\rho_m}-n_kq_k\frac{m_im_j}{q_iq_j\rho_m}(\frac{q_i}{m_i} 
+\frac{q_j}{m_j}-\frac{q_k}{m_k})\right)
({\bf E}+ {\bf v_k}\times {\bf B})
-\frac{m_im_j}{q_iq_j\rho_m}\frac{\partial {\bf J}}{\partial t} \\ \nonumber
+\frac{m_im_j}{q_iq_j\rho_m}(\frac{q_j}{m_j}+\frac{q_i}{m_i})
({\bf J}\times {\bf B})  
 -(\frac{m_im_j}{q_iq_j\rho_m})\bigg(\frac{q_i}{m_i}\nabla{p_i}+
 \frac{q_j}{m_j}\nabla{p_j}+\frac{q_k}{m_k}\nabla{p_k}\bigg) 
 -(\frac{m_im_j}{q_iq_j\rho_m} )\bigg[{\bf J}(\nu_{ik}+\nu_{kj}+\nu_{ji}) \\ \nonumber
 +\rho_m{\bf V}(\frac{q_i}{m_i}\nu_{ij} +\frac{q_j}{m_j}\nu_{jk}
 +\frac{q_k}{m_k}\nu_{ki})
 -\nu_{ki}\bigg((n_kq_k+\frac{n_im_iq_k}{m_k}){\bf v}_i 
 + (\frac{n_km_k}{n_im_i}n_jq_j +\frac{n_jm_jq_k}{m_k}){\bf v}_j \\ \nonumber
  -(\frac{n_km_k}{n_im_i}n_jq_j){\bf v}_k\bigg)
  - \nu_{ij}\bigg((-\frac{n_im_i}{n_jm_j}n_kq_k){\bf v}_i 
  +(n_iq_i+\frac{n_jm_jq_i}{m_i}) {\bf v}_j
 +(\frac{n_im_i}{n_jm_j}n_kq_k +\frac{n_km_kq_i}{m_i}){\bf v}_k\bigg) \\ \nonumber 
 -\nu_{jk}\bigg((\frac{n_jm_j}{n_km_k}n_iq_i 
 +\frac{n_im_iq_j}{m_j}){\bf v}_i -(\frac{n_jm_j}{n_km_k}n_iq_i){\bf v}_j
 +(n_jq_j+\frac{n_km_k q_j}{m_k}){\bf v}_k\bigg) \bigg].
\end{eqnarray}

This is an extended form of generalized Ohm's law, which relates the 
electric field ${\bf E}$ to the global variables ${\bf V}$ and ${\bf J}$. 
% In right hand side, the first term arises from the inertial flows, 
% the second term arises because of hall current across the magnetic 
% field, the third term is sue to pressure variation, and the 
% last term is due to Coulomb collisions which 
% gives resistivity effects in the flow system.
This equation can readily reduces to the well-known generalized 
Ohm's law and simplified Ohm's law for general two fluid plasma 
if the contributions associated with the third 
component ($n_k, q_k, {\bf v_k}$) is removed from the above equation 
in the MHD limit~\cite{doi:10.1029/1998JA900065,doi:10.1002/ctpp.19650050103}. 
\end{widetext}
%%%%%%%%%%%%%%%%%%%%%%%%%%%%%%%%%%%%%%%%%%%%%%%%%%%%%%%%%%%%%%%%%%%%%
%   \bibliography{../bibliography_all_2018_fusion.bib}% Produces the bibliography via BibTeX.

% \end{thebibliography}
%%%%%%%%%%%%%%%%%%%%%%%%%%%%%%%%%%%%%%%%%%%%%%%%%%%%%%%%%%%%%%%%%%%%%%%
\end{document}